\UseRawInputEncoding
\documentclass[journal=jacsat,manuscript=article]{achemso}

\usepackage{graphicx} % Required for inserting images
\usepackage{amsmath}
\usepackage{float}
\usepackage{xcolor}
\usepackage{dirtytalk}
\usepackage{soul}

\author{Gracie Chaney}
\affiliation{Université Grenoble Alpes, CEA, LITEN, 17 rue des Martyrs, 38054 Grenoble, France}

\author{Andrey Golov}
\affiliation{Centre for Cooperative Research on Alternative Energies (CIC energiGUNE), Basque Research and Technology Alliance (BRTA), Alava Technology Park, Albert Einstein 48, 01510 Vitoria‐Gasteiz, Spain}

\author{Ambroise van Roekeghem}
\affiliation{Université Grenoble Alpes, CEA, LITEN, 17 rue des Martyrs, 38054 Grenoble, France}

\author{Javier Carrasco}
\affiliation{Centre for Cooperative Research on Alternative Energies (CIC energiGUNE), Basque Research and Technology Alliance (BRTA), Alava Technology Park, Albert Einstein 48, 01510 Vitoria‐Gasteiz, Spain}
\alsoaffiliation{Ikerbasque, Basque Foundation for Science, Plaza Euskadi 5, 48009 Bilbao, Spain}

\author{Natalio Mingo}
\affiliation{Université Grenoble Alpes, CEA, LITEN, 17 rue des Martyrs, 38054 Grenoble, France}

\title{
Two-step growth mechanism of the solid electrolyte interphase in argyrodyte/Li-metal contacts 
}

\begin{document}

\begin{abstract}
The structure and growth of the Solid Electrolyte Interphase (SEI) region between an electrolyte and an electrode is one of the most fundamental, yet less-well understood phenomena in solid-state batteries. We present a parameter-free atomistic simulation of the SEI growth for one of
the currently promising solid electrolytes (Li$_6$PS$_5$Cl), based on \textit{ab initio} trained machine learning (ML) interatomic potentials, for over 30,000 atoms during 10 ns, well-beyond the capabilities of conventional MD. This unveils a two-step growth mechanism: Li-argyrodite chemical reaction leading to the formation of an amorphous phase, followed by a kinetically slower crystallization of the reaction products into a 5Li$_2$S·Li$_3$P·LiCl solid solution. The simulation results support the recent, experimentally founded hypothesis of an indirect pathway of electrolyte reduction.  These findings shed light on the intricate processes governing SEI evolution, providing a valuable foundation for the design and optimization of next-generation solid-state batteries.
\end{abstract}
\begin{frame}{TOC graphic}
\begin{figure}%[H]
\begin{center}
\includegraphics[width=7.5cm]{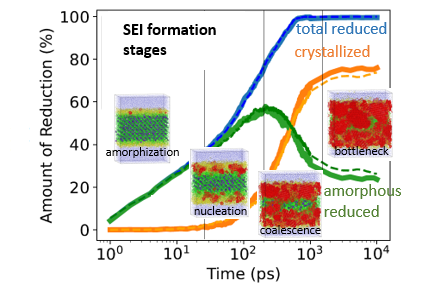}
\caption*{TOC graphic}
\label{TOC}
\end{center}
\end{figure}
\end{frame}

\maketitle
%TC:endignore
Li-metal anode all-solid-state batteries (ASSBs) may replace current Li-ion batteries, since they have higher theoretical energy densities,\cite{wang_batteryComparison_2023} 
\cite{zhao_designing_2020} 
 are potentially safer than their liquid electrolyte counterparts
\cite{cheng_recent_2019} 
and may also achieve 
longer cycling life and high power densities. 
\cite{janek_solid_2016} 
Sulfide argyrodite electrolytes, such as Li$_{6}$PS$_{5}$Cl (LPSC),  are particularly promising for ASSB's due to their combination of high ionic conductivities and mechanical softness.\cite{chen_sulfide_2018,park_design_2018,zhang_new_2018,kudu_review_2018,wang_design_2015} 
However LPSC is %
highly reactive when in contact with a Li-metal anode.\cite{ceder_sulfides_2020}  During the reaction of LPSC with the anode, an  
SEI 
layer 
develops from the reduction of the thiophosphate groups. Although an SEI can hinder the rate of charge and discharge, its presence is sometimes able to protect the electrolyte from further degradation, and to increase the electrochemical stability window. This is crucial for the practical application of argyrodites in ASSB's. Understanding the physics behind the growth of the SEI in Li-metal/argyrodite contacts is therefore a priority. Experimental evidence for how this SEI forms in LPSC is however limited, and a detailed description of this process at the atomic level is currently a very sought after goal.\cite{wenzel_interfacial_2018,schwietert_clarifying_2020} 

Ever since the concept of SEI was proposed in 1979,\cite{peled_advanced_1997} many modeling efforts have attempted to decipher the microscopic mechanisms behind the SEI's initial formation stages and long term growth. Despite these models being mutually incompatible, most of them are able to match the experimentally reported SEI thickness evolution, due to their use of adjustable parameters.(see Single \textit{et al.}\cite{single_dynamics_2016,single_identifying_2018,single_theory-based_2021}). Thus, it is urgent to be able 
to perform 
parameter-free, 
\textit{ab initio} simulations of SEI growth. 
In the specific case of solid-state electrolytes, 
recent progress has taken place by combining bulk \textit{ab initio} calculations with analytical expressions and differential equations, in order to obtain the potential profile along the SEI and elucidate the main ion transport mechanisms involved in its growth.\cite{swift_first-principles_2019,swift_modeling_2021} However, fully atomistic modeling studies are still scarce. 
\textit{Ab initio} molecular dynamics (AIMD) modeling has been performed.\cite{golov_2021,cheng_quantum_2017,camacho-forero_exploring_2018,wang_ionic_2020}  
However,  
AIMD is a resource-intensive method,  
where simulations are 
limited to time scales of hundreds of picoseconds, and the interface models include fewer than a thousand atoms, which seem insufficient to capture a realistic picture of the SEI growth.  
This problem can be circumvented by the use of 
\textit{ab initio} trained 
machine learning interatomic potentials (MLIPs).\cite{Behler_2007, Behler_2015, Behler_2016, MLIP_botu_2017, MLIP_deringer_2019, MLIP_gokcan_2022, friederich_machine-learned_2021}

\begin{figure}%[H]
\begin{center}
\includegraphics[width=8.0cm]{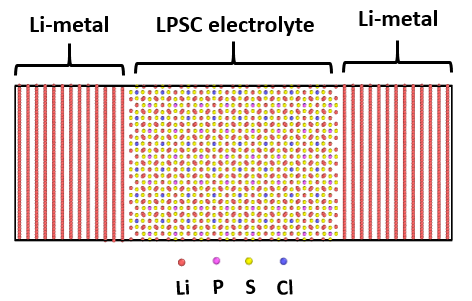}
\caption{Li(110)/LPSC(110) interface model I.
}
\label{crystal-orientations}
\end{center}
\end{figure}

Thus, the urgent need to understand the SEI growth in Li-metal/solid electrolyte contacts at the atomic level, and the limitations inherent to AIMD in previous simulations,
have prompted us to carry out large scale parameter-free ML-MD simulations on over 30,000 atoms for tens of nanoseconds. The results unveil a rich and complex mechanism, inaccessible to shorter, smaller sized simulations.

The paragraphs below describe 
our use of moment tensor potentials (MTPs) with the MLIP-2 code\cite{Novikov_2021} to simulate SEI formation between a Li-metal anode and an LPSC electrolyte at various temperatures, internal pressures, and initial velocities. We discover a two-step growth process consisting of an initial electrolyte reduction followed by gradual crystallization %
of the reaction product.  
The value of the exponent in the crystallization's time dependence %
appears to be influenced by the MD temperature, suggesting 
a qualitative change in %
the driving mechanism of crystallization.  Finally, we found that  crystallization of the reaction product impedes the movement of Li from the anode, 
which can slow down the rate of further  
reduction.

\begin{figure}[b!]
\begin{center}
\includegraphics[width=8.8cm]{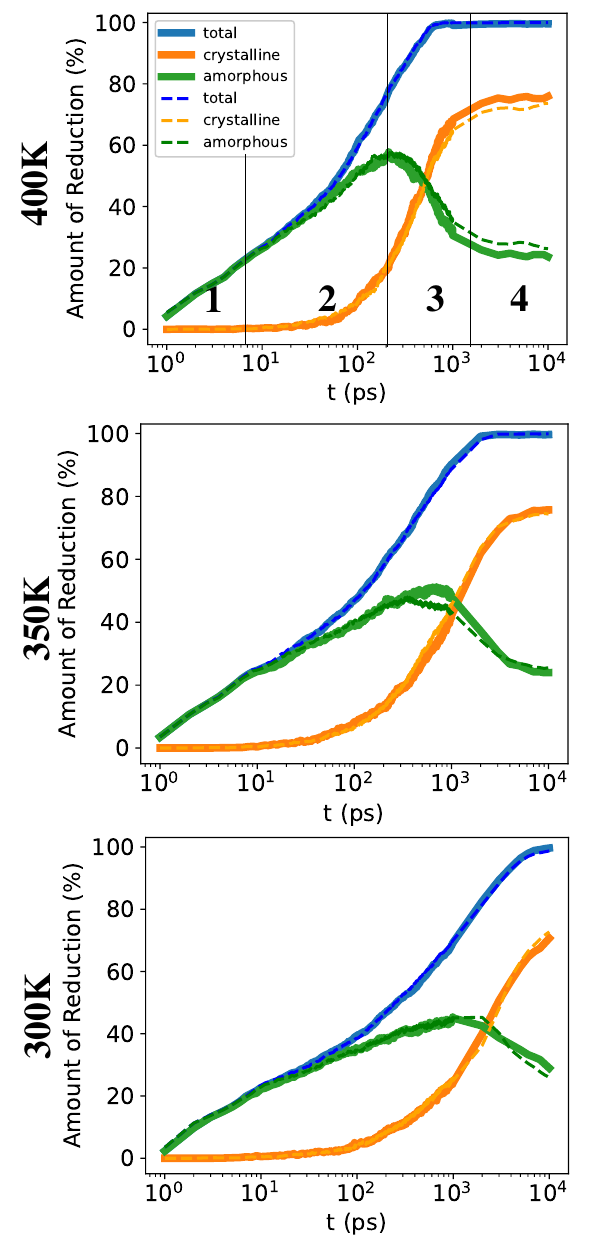}
\caption{
Percentage of reduced solid electrolyte vs time, model II. The thick and thin curves correspond to two independent MD simulations with different initial velocities.}
\label{phases-largeSC}
\end{center}
\end{figure}

\begin{figure}[b!]
\begin{center}
\includegraphics[width=\textwidth]{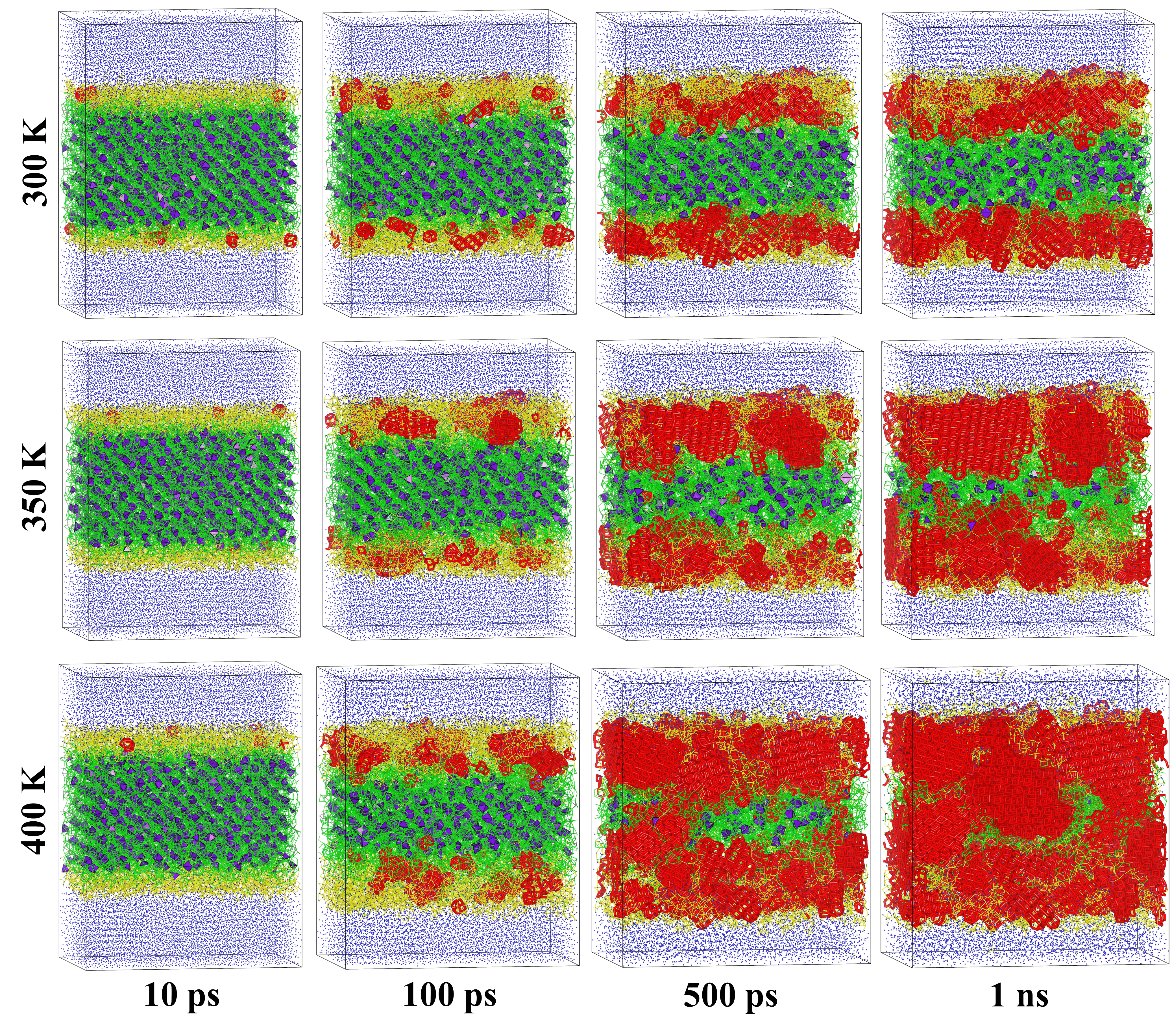}
\caption{ 
Evolution of the SEI. Red: crystalline regions. Yellow: amorphous reduced regions. Blue dots: Li-metal. Purple polyhedra and green lines correspond to thiophosphate groups and Li-Cl, Li-S bonds, respectively, in the original argyrodite structure. The snapshots correspond to the model II interface for various temperatures at a pressure of 1 bar (seed no. 1 for the initial velocities.) For the sake of clarity, the snapshots at 10 ns, showing full crystallization, are included in the SI rather than in this figure.
}
\label{bottle-neck}
\end{center}
\end{figure}

We build two Li(110)/LPSC(110) interface models of different sizes following the methodology described in Ref.~\cite{golov_2021}. The numbers in parentheses denote the crystallographic directions of Li and LPSC planes at the interface. Model I includes LPSC and metal Li slabs with 288 formula units and 4212 atoms, respectively (7956 atoms in total) (Fig.~\ref{crystal-orientations}). Model II includes 1152 LPSC formula units and 16848 atoms of metal Li (31824 atoms in total). The interface surface areas of models I and II are 1345~\AA$^2$ and 5380~\AA$^2$, respectively.
Direct visualization of the MD trajectories shows the SEI to be a mixture of amorphous and crystalline regions, the latter increasingly dominating over the former as time goes by (Figs.~\ref{phases-largeSC} and \ref{bottle-neck}). To identify and quantify the extent of the crystalline regions, we employ a methodology developed in Ref.~\cite{golov_unveiling_2023} (see below).
In turn, we define the amorphous region as the part of the reduced electrolyte that is not crystalline. 
Fig.~\ref{phases-largeSC} displays the evolution of these phases over time, at three different temperatures, from two separate simulations per temperature, where the velocities at the first time step where initialized from two different random seeds (see SI). 
The curves in Fig.~\ref{phases-largeSC} 
clearly show the two basic steps leading to SEI growth: 
an initial reduction of argyrodite by metal Li results in the formation of the amorphous phase, followed by the crystallization of the reaction product. This observation is in line with the earlier hypothesis that this solid-state reduction proceeds indirectly via a metastable overlithiation, which later crystallizes in the stable products.\cite{schwietert_clarifying_2020,schwietert_first-principles_2021}. 

As previously 
shown using AIMD, our results also find that the crystalline products are not a phase separated mixture of LiCl, Li$_2$S, and Li$_3$P, but a 5Li$_2$S·Li$_3$P·LiCl solid solution, which corresponds to the antifluorite structure type.\cite{golov_unveiling_2023} 
This is not incompatible with experimental evidence: XPS measurements have been employed to identify
the three products,\cite{wenzel_interfacial_2018} however this experimental method,
based on characterizing electronic binding energies of individual chemical species, faces difficulties in distinguishing between phase-separated domains, or a singular solid
solution, where the S sublattice comprises individual substitutions by P and Cl. Indeed, in this study part of the signal was attributed to unknown reduced P species, and the possibility of having a solid solution forming rather than separate phases for the different thermodynamically stable products has also been postulated.\cite{schwietert_first-principles_2021}

In terms of microstructure, the simulation allows us to identify four different reaction stages. Let's take, for example, the 400~K, 1~bar results of Figs.~\ref{phases-largeSC} and \ref{bottle-neck}:

(1) Within the first 5~ps of the simulations, the reaction between Li and the electrolyte leads to the formation of a thin layer of amorphous product. 

(2) Starting from 5~ps, the first crystalline nuclei appear in the amorphous matrix. The reaction continues with the growth of amorphous and crystalline regions. As discussed below, electrolyte conversion has a logarithmic dependence on time, while SEI crystallinity  
has a square root (at 300~K) or linear (at 400~K) dependence. Thus the amorphous phase dominates over the crystalline at the initial stages of the reaction.  

(3) At  
165~ps, 
electrolyte conversion reaches 71\%. This point corresponds to the maximal amount of amorphous phase within reduced electrolytes. This is also when the so far separate crystalline domains start to coalesce. Subsequent growth of crystalline nuclei with a limited amount of unreacted electrolyte will lead to a decrease in the percentage of amorphous product, and a drastic slowdown in the amount of Li diffusing into the electrolyte.
This slowdown might be related to the decreased Li diffusivity in the crystalline region compared to that in the amorphous one, and also to the fact that as time progresses there remains less of the original LPSC left to react with the diffusing Li. 

(4) At 1 ns, all electrolyte is reduced and the crystal nuclei reach their maximum size. A further slight increase in crystallinity is due to changes in the structure of grain boundaries. We cannot expect 100\% crystallinity even with longer MD simulations due to the presence of defects and grain boundaries.%}
 
Fig.~\ref{phases-largeSC}, also shows that as temperature decreases, the system takes longer to reach its long-term growth regime, which can be attributed to the sluggish atomic kinetics.  The cases with 1 bar of internal pressure are rather similar to those with 1 kbar of internal pressure; 
future studies should explore larger internal pressures for a comprehensive understanding.  

 An important finding is that the amount of reduced electrolyte in the simulation increases over time $t$ roughly as $\log(t)$ (Fig.~\ref{phases-largeSC}). This behavior can arise if the rate of Li inflow into the electrolyte is not limited by diffusion, but by the difference in Li chemical potentials, $\Delta \eta = \mu_{Li}^A-\mu_{Li}^E$ between the metal anode and the electrolyte. According to the Butler-Volmer approximation (assuming a symmetry factor $\alpha=0.5$), 
$J\propto e^{\frac{\Delta\eta}{2 k_BT}}- e^{-{\frac{\Delta\eta}{2 k_BT}}}$.\cite{van_der_ven_rechargeable_2020} 
The difference in chemical potentials 
progressively decreases as the 
electrolyte’s Li 
concentration, $n$, 
increases, and it becomes zero when the electrolyte is completely reduced, at 
$n=n_\text{r}$. 
As a first approximation we can then write
\begin{eqnarray}
    {\frac{1}{2}}{\frac{\Delta\eta}{k_B T}} = \gamma (n-n_r),
\end{eqnarray}
which for $n<<n_r$ yields
\begin{eqnarray}
    J\propto e^{-\gamma(n - n_\text{r})}\equiv 
    A e ^{-\gamma n},
\end{eqnarray}
where $A=e^{\gamma n_\text{r}}$.
Since the amount of reduced electrolyte is proportional to $n$, we have 
\begin{eqnarray}
    {\frac{dn}{dt}}= J\propto e^{-n}.
\end{eqnarray}

Therefore $dt\propto {\frac{dn}{e^{-n}}} \Rightarrow t\propto e^n$, 
which implies $n\propto \log(t)$.

 The 
 intuitive picture behind this logarithmic law is one where Li atoms roam unimpeded through the electrolyte, and the only limit to their flow comes from the dwindling chemical potential difference between the two reacting materials. The $\log(t)$ behavior also suggests that the electrolyte becomes reduced via a continuous lithiation process through which there is no sharp structural phase transition. It is only after being reduced, that the (by then amorphous) structure crystallizes, with an associated decrease of Li mobility. (Logarithmic growth laws are not the most common ones in physics. Another example of logarithmic behavior is the time dependence of p-MOSFET degradation, which however has a different origin.\cite{wang_explanation_1991})

\begin{figure}[t!]
\begin{center}
\includegraphics[width=8.5cm]{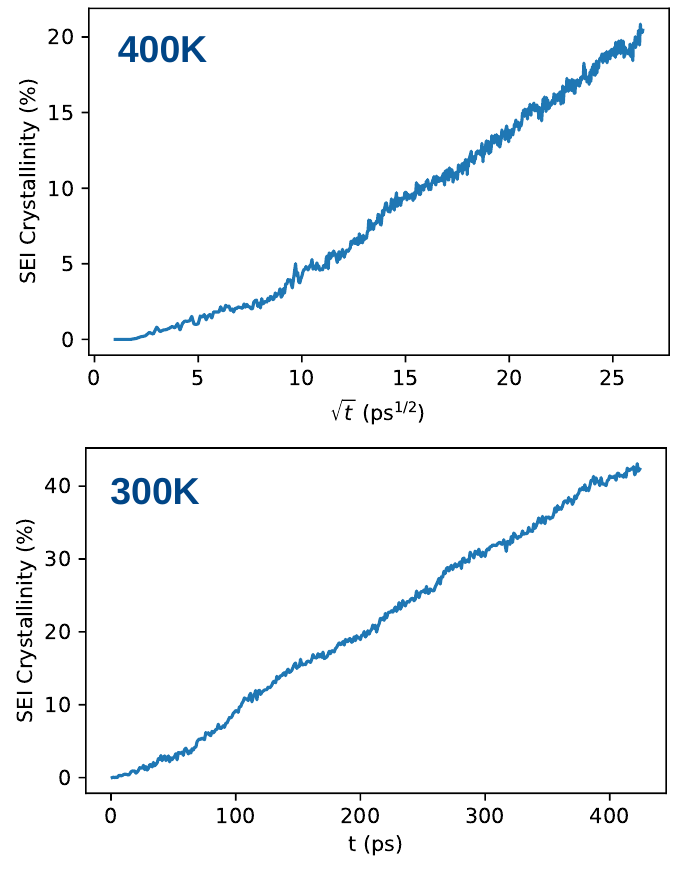}
\caption{Amount of polycrystalline reduced electrolyte, given as a percentage of the total possible reduction, as a function of time, for the model II interface at 1 bar of internal pressure. Notice the different time scales.}
\label{crystal-comparison}
\end{center}
\end{figure}

After the initial reduction of the  
electrolyte, %
crystallization occurs. 
The crystallization time dependence is very different from that of the amorphous reduction, and its functional form depends on temperature (Fig. \ref{crystal-comparison}).  For 300~K, crystallization begins at approximately %2 ps 
10~ps %
and grows with a square-root of time dependence.  Such a trend is common for diffusion-limited processes.\cite{chen_annealing_2014,hay_secondary_2018,louat_theory_1974}  
For 400~K, however, crystallization acquires a linear time dependence, which is much more common for interface-limited processes.\cite{chason_effect_1991}  

As mentioned earlier the rates of reduction and crystal growth appear to plateau in Fig.~\ref{phases-largeSC}. It is tempting to associate this to the nearly complete reduction of the electrolyte. Nevertheless, certain simulations for the smaller system (specifically, the 1~kbar, seed~1 case of Fig. S1) show a reaction that stalls before all of the available electrolyte is reduced. This could be attributed to a negative-feedback effect in which Li-diffusion initially promotes crystallization by reducing the electrolyte, %
and then the growing crystals impede further Li movement within the electrolyte.  It appears from Fig. \ref{bottle-neck} that separate 
crystalline nuclei %
(shown in red) begin near the anode layers, before extending along all directions in three-dimensional space.  Eventually, the crystal regions merge.  At first, there are some pores left for the input Li to enter the electrolyte, but by 1 ns in Fig. \ref{bottle-neck}, the crystal regions block nearly all incoming Li (see also section~S4 of the SI).   

Interestingly, the effect of different seeds on the initial velocities is marked only for the smaller systems, but becomes minor on the largest (31,824 atom) interface models.  
We believe this is due to the fact that crystallization proceeds akin to a percolation process. Throughout reaction stages 2, 3 and 4 (see Fig.~\ref{phases-largeSC}), the crystalline domains start nucleating (2), coalescing (3), and finally merging into a continuous polycrystalline layer (4) that splits the electrode off the rest of the electrolyte, at which the reaction nearly stalls due to lack of Li inflow. The start of stage 4 can be viewed as the percolation threshold, at which there are no more amorphous regions left that continuously link electrode and remaining electrolyte. It is known in percolation theory that the experimental or simulated value of a system's percolation threshold becomes more reliable as the system size grows. This explains why the 
model I systems still present rather different electrolyte reduction and SEI crystallinity curves (Fig. S1), depending on the seed, whereas the 
model II ones %
have a good match of the curves up to 10~ns (Fig.~\ref{phases-largeSC}).

The initial stages of SEI formation just described are a fast process taking place at the timescale of nanoseconds (ns), after which the slower, long-term growth begins, limited by the less ion-conductive 
5Li$_2$S·Li$_3$P·LiCl solid solution polycrystalline layer. 
Directly observing the initial reaction stages poses a challenge. 
Perhaps nano-indentation techniques, in combination with in operando transmission electron microscopy, might enable real time imaging of the SEI time evolution at the atomistic level, in a similar way as the reorganization of metal chains during break-junction experiments was imaged some 25 years ago.\cite{ohnishi_quantized_1998}

The results presented are for unbiased electrolyte/electrode contacts. A welcome future development would be the ability to perform similar atomistic simulations in the presence of applied external electric fields and perhaps finite electron currents. This will require the use of new methods, including fourth-generation machine learning interatomic potentials under bias, and possibly involving non-equilibrium electron transport approaches.

In conclusion, we 
have deciphered the evolution of the atomic structure during the early stages of SEI growth at a Li-metal/LPSC interface.  Machine learning-enhanced MD has enabled us to simulate systems up to 36 times larger than those previously modeled by AIMD, for trajectories up to 100 times longer.\cite{golov_2021}  This much larger timescale unveils previously unknown facts: 

1) SEI formation is a two step process: Li-LPSC redox reaction leading to the formation of the amorphous phase, followed by the crystallization of the reaction products into a polycrystalline solid solution.  This aligns with Schwietert \textit{et al.'s} hypothesis of an indirect pathway.\cite{schwietert_clarifying_2020}    

2) The redox reaction follows a logarithmic time dependence, suggesting that it is driven by a chemical potential difference. In turn, crystallization increases either linearly, or as a square-root of time, depending on temperature. This suggests interface- and diffusion-limited processes, respectively.   

3) Over the 10 ns simulation we observe the onset of four distinct reaction stages: formation of the amorphous product, crystal nucleation, crystal coalescence, and reaction stalling. After the onset of the last stage, growth proceeds at orders of magnitude slower rate. The growth curves become fairly consistently reproducible, irrespective of initial velocities, when the system size exceeds 30,000 atoms.

By employing predictive atomistic simulations we can surpass inconclusive models of SEI growth based on adjustable parameters. The outcomes and insights from this study, focusing on the LPSC/Li-metal contact, serve as an illustration of the potential of this approach for various other solid electrolyte systems currently under investigation. Further developments in the field of ASSBs can highly benefit from large scale parameter-free atomistic simulations akin to the ones showcased in this work.

GC, AVR and NM thank T. Ayadi, F. Bruneval, and M. Nastar for helpful discussions. Project funded by CEA through program FOCUS-batteries.

\bibliography{bibliography.bib}

\end{document}